\documentclass[aps,prb,twocolumn,superscriptaddress]{revtex4-1}

\usepackage{graphicx,epstopdf,siunitx,braket,xspace,amssymb,dsfont,bm,todonotes, natbib,amsmath}
\setcitestyle{super, sort} 
\begin{document}
	
	\title{Quadrupolar and anisotropy effects on dephasing in two-electron spin qubits in GaAs}
	
	\author{Tim Botzem}
	\author{Robert P. G. McNeil}
	\affiliation{JARA-Institute for Quantum Information, RWTH Aachen University, D-52074 Aachen, Germany}
	\author{Dieter Schuh}
	\affiliation{Institut f\"ur Experimentelle und Angewandte Physik, Universit\"at Regensburg, D-93040 Regensburg, Germany\\}
	\author{Dominique Bougeard}
	\affiliation{Institut f\"ur Experimentelle und Angewandte Physik, Universit\"at Regensburg, D-93040 Regensburg, Germany\\}
	\author{Hendrik Bluhm}
	\affiliation{JARA-Institute for Quantum Information, RWTH Aachen University, D-52074 Aachen, Germany}
	\date{\today}
	

	\maketitle
\textbf{
Understanding the decoherence of electron spins in semiconductors due to their interaction with nuclear spins is of fundamental interest as they realize the central spin model\cite{chen_semiclassical_2007} and of practical importance for using electron spins as qubits.
Interesting effects arise from the quadrupolar interaction of nuclear spins with electric field gradients, which have been shown to suppress diffusive nuclear spin dynamics \cite{welander_influence_2014}. One might thus expect them to enhance electron spin coherence \cite{sinitsyn_role_2012}. 
Here we show experimentally that for gate-defined GaAs quantum dots, quadrupolar broadening of the nuclear Larmor precession can also reduce electron spin coherence due to faster decorrelation of transverse nuclear fields.
However, this effect can be eliminated for appropriate field directions. Furthermore, we observe an additional modulation of spin coherence that can be attributed to an anisotropic electronic $\boldsymbol{g}$-tensor.
These results complete our understanding of dephasing in gated quantum dots and point to mitigation strategies. They may also help to unravel unexplained behaviour in related systems such as self-assembled quantum dots\cite{press_ultrafast_2010} and III-V nanowires\cite{nadj-perge_spin-orbit_2010}.
}

\begin{figure}[!ht]
	\centering
	\includegraphics[trim = 0cm 0cm 0cm 0cm,width=\textwidth/2]{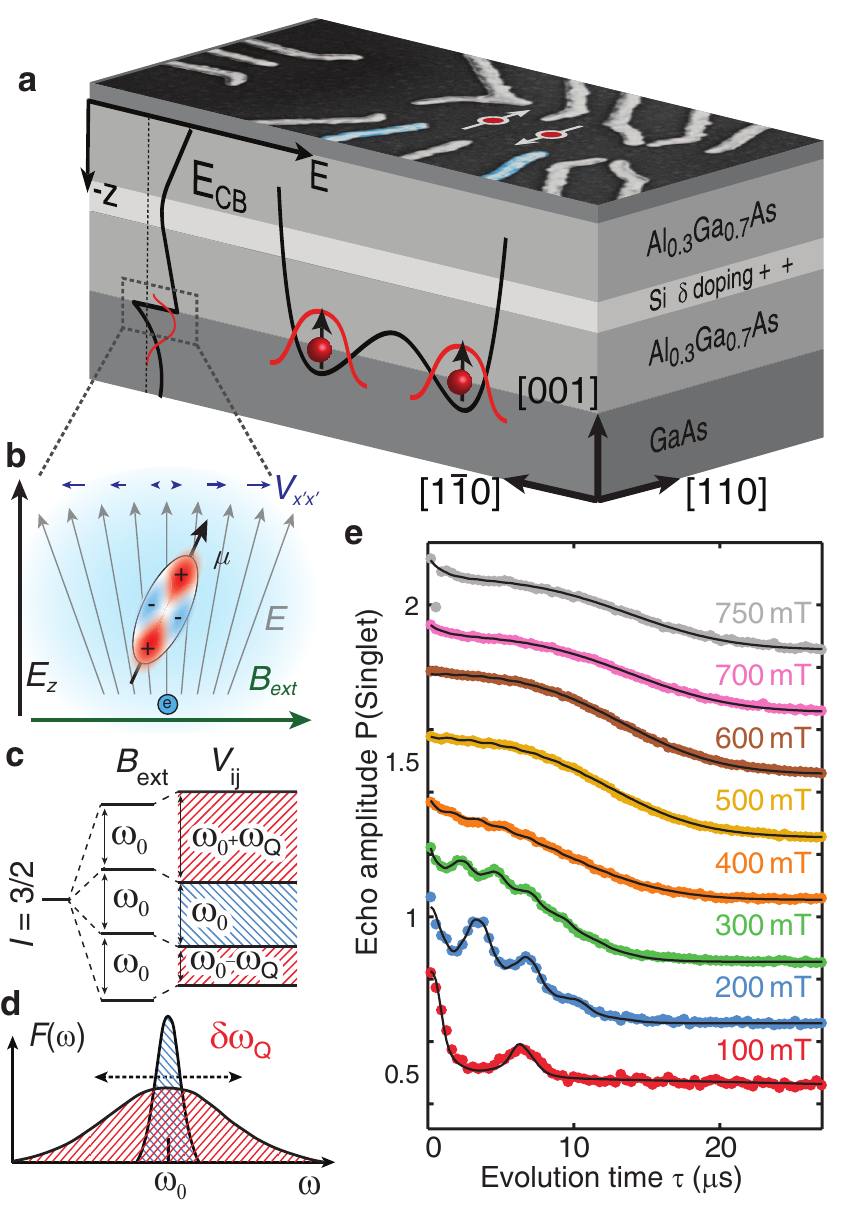}
	\caption{\textbf{Device Layout and quadrupole boadening} \textbf{a}, Gates used for pulsed qubit control are depicted in blue; the energy of the conduction band edge $E_{\mathrm{CB}}$ is shown on the left. \textbf{b}, Nuclear spins 3/2 in the proximity of the quantum dot experience quadrupolar coupling to electric field gradients $V_{x'x'}$ induced by crystal distortion due to the electric field of the triangular quantum well. \textbf{c}, While the center transition stays unchanged, the satellite transitions, distorted by the electron's own charge, exhibit a quadrupolar shift by $\omega_Q$. \textbf{d}, The resulting frequency distribution $F(\omega)$ consists of two Gaussians with different variances. \textbf{e}, Echo amplitude for magnetic fields along the $[110]$ axis, showing oscillations with the relative Larmor frequencies of the three nuclear spins. A semi-classical model (solid line) is used to fit the data (dots, offset for clarity).}
	\label{fig:echoZ}	
\end{figure}
Electron spin qubits in GaAs quantum dots have played a central role in demonstrating the key operations of semiconductor spin qubits \cite{petta_coherent_2005, barthel_rapid_2009, foletti_universal_2009, shulman_demonstration_2012}.
A prominent and often dominant dephasing mechanism in these  devices as well as other semiconductor spin qubits \cite{press_ultrafast_2010, greilich_mode_2006} is the interaction of the electron spin with $10^4-10^6$ nuclear spins of the host lattice.
While the fundamentals of this interaction have been studied quite extensively\cite{cywinski_pure_2009, cywinski_electron_2009, witzel_quantum_2006, yao_theory_2006}, and theory and experiments are in reasonable agreement \cite{bluhm_dephasing_2011}, theory predicts a potential for much longer dephasing times \cite{lee_universal_2008} than observed so far and it remains an open question as to what ultimately limits electron spin coherence. 
Remarkable progress has also been made in eliminating dephasing from nuclear spins by using Si-based systems\cite{zwanenburg_silicon_2013} that can be isotopically purified, but this route is not open for III-V semiconductor systems, where all isotopes carry nuclear spin. Nevertheless, the latter remain of practical interest because of their lower effective mass, single conduction band valley and potential for optical coupling.
Recently the role of quadrupolar coupling of nuclear spins to electric field gradients (EFGs) from charged impurities or strain has been investigated both experimentally and theoretically\cite{welander_influence_2014, sinitsyn_role_2012, munsch_manipulation_2014, chekhovich_suppression_2015, maletinsky_dynamics_2007}, but its influence on electron spin coherence is still unclear.
\begin{figure*}
	\centering
	\includegraphics[trim = 0cm 0cm 0cm 0cm, width=\textwidth]{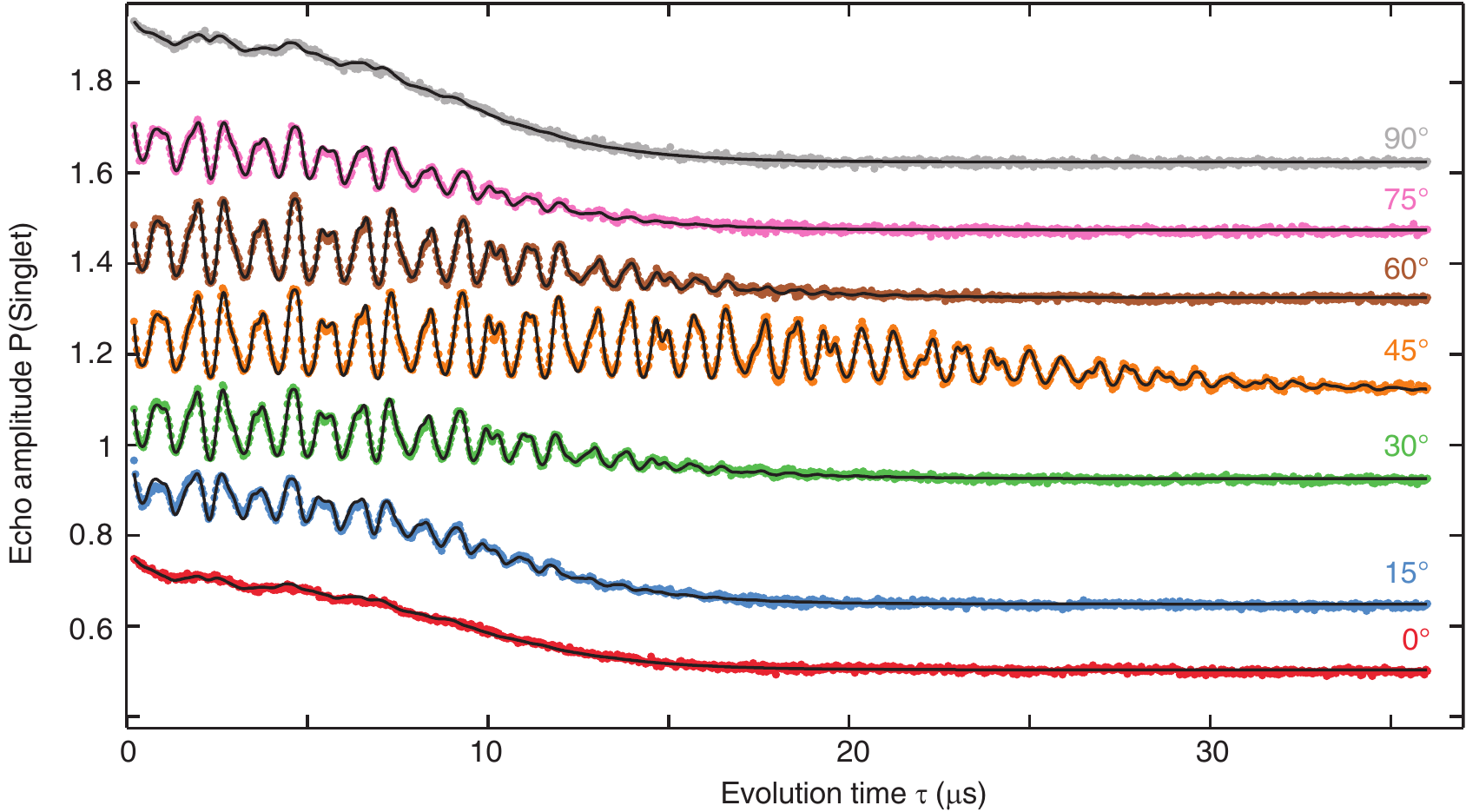}
	\caption{\textbf{$\mathbf{B}$-field direction dependence} Echo amplitude at $300\,$mT as a function of separation time for different in-plane magnetic field directions $\theta$, with $0^\circ$ corresponding to  the $[110]$ direction. Curves are offset for clarity. At $45^{\circ}$, parallel to the crystallographic [100] axis, the coherence time is enhanced as quadrupolar couplings are suppressed. When rotating the field a $g$-factor anisotropy leads to oscillations, associated with the three different nuclear Larmor frequencies. A semi-classical model (solid line) is used to fit the data (dots).}
	\label{fig:echoY}	
\end{figure*}
\newline
In contrast to the expected enhancement of coherence due to quadrupolar suppression of nuclear spin flip-flops, we find that Hahn echo coherence improves when the magnetic field is rotated to minimize quadrupolar broadening of nuclear levels. In addition, we find a complex pattern of collapses and revivals of the echo signal unless the magnetic field is aligned with specific crystal axes, which we explain with an anisotropic $g$-tensor causing a coupling of the nuclear Larmor precession to the electron spin.

The qubit studied here is a two-electron spin qubit\cite{levy_universal_2002,petta_coherent_2005},  using the $m_{z}=0$ subspace $S = \left(\ket{\uparrow \downarrow} - \ket{\downarrow\uparrow}\right)/\sqrt{2}$ and $T_{0}= \left(\ket{\uparrow \downarrow} + \ket{\downarrow\uparrow}\right)/\sqrt{2}$ of two electron spins. These electrons are confined in a GaAs double quantum dot formed by electrostatic gating (Fig.$\,$\ref{fig:echoZ}a) of a two-dimensional electron gas (2DEG). The effects explored in this work apply equally to single electron spins.  

A random configuration of the nuclear spins introduces an effective magnetic field of a few mT, the Overhauser field, whose dynamics cause qubit dephasing. Hahn echo measurements that eliminate dephasing from slow fluctuations allow studying these dynamics, as they become the dominant dephasing mechanism.
\begin{figure*}
\centering
	\includegraphics[trim = 0cm 0cm 0cm 0cm, width=\textwidth]{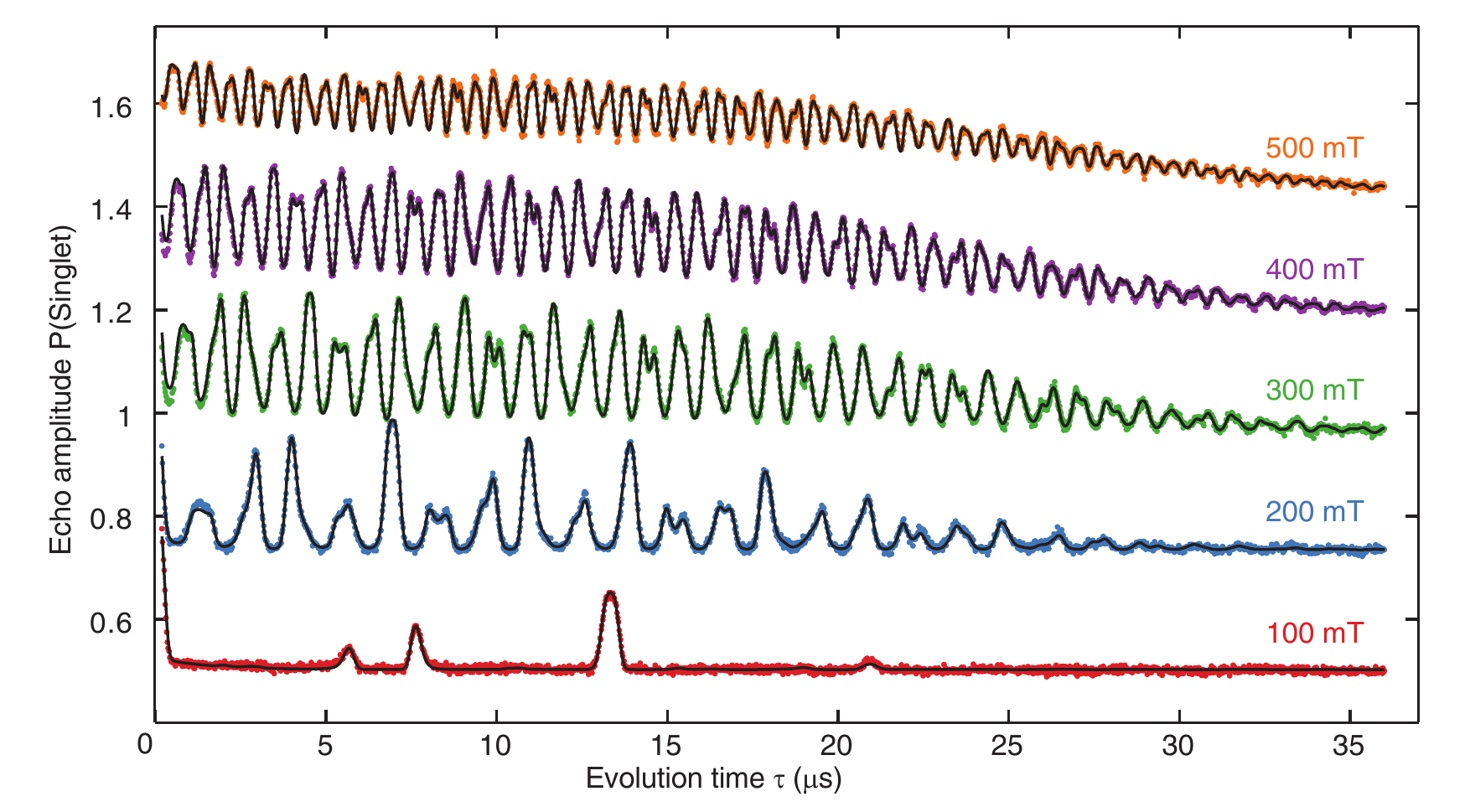}
	\caption{\textbf{$\mathbf{B}$-field magnitude dependence} Echo amplitude for magnetic field magnitudes along the [100] axis. A $g$-factor anisotropy causing different quantization axes for electron and nuclei spins leads to oscillations with the three nuclear Larmor frequencies. For small magnetic fields the echo signal is strongly suppressed in the first hundreds of nanoseconds, but revives at later times. A semi-classical model (solid line) is used to fit the data (dots).}
	\label{fig:echoXZ}	
	\end{figure*}
We follow the experimental procedure from Ref.\,\onlinecite{bluhm_dephasing_2011} (see also Methods), implementing the required $\pi$-pulse to invert the state of the qubit halfway through the evolution time $\tau$ using the exchange interaction between the two spins.
Fig.\,\ref{fig:echoZ}c shows the spin echo signals as a function of separation time for magnetic fields aligned along the $[110]$ crystal axis. (Note that we experimentally cannot distinguish between the $[110]$ and $[1\overline 1 0]$ axes, but refer to the direction parallel to the dot connection line as the $[1\overline 1 0]$ axis throughout the paper for ease of reading.) Similar results to Refs.\,\onlinecite{bluhm_dephasing_2011, medford_scaling_2012} are obtained, but with approximately a factor two shorter coherence times (see Supplementary Information).
At fields below $500\,$mT, a second order coupling to the oscillating, transverse nuclear field (i.e., its component perpendicular to the external field) leads to periodic collapses and revivals of the echo amplitude\cite{witzel_quantum_2006, yao_theory_2006, bluhm_dephasing_2011}. Revivals occur at times corresponding to the periods of the relative Larmor precession of the three species $^{69}$Ga, $^{71}$Ga and $^{75}$As. The overall envelope decay can be modeled by assuming a phenomenological broadening $\delta B$ of the Larmor frequencies that causes fluctuations of the transverse hyperfine field of each species. 

While such a broadening is expected from dipolar interaction between nuclei, fitting the current and earlier\cite{bluhm_dephasing_2011} data requires a value of $\delta B = 1.4\,$mT and $\delta B = 0.3\,$mT respectively, at least a factor three larger than the intrinsic dipolar nuclear linewidth of $0.1\,$mT obtained from NMR measurements in pure GaAs\cite{sundfors_exchange_1969}. More direct measurements of the nuclear dynamics based on correlation of rapid single shot measurements \cite{dickel_characterization_2014} are consistent with these values.

NMR experiments on GaAs samples with impurities revealed similar excess line broadening  which was found to depend on the field direction and explained by quadrupolar effects\cite{sundfors_exchange_1969, hester_nuclear-magnetic-resonance_1974}. Electric fields from charged impurities, strain or the triangular quantum well, used here to confine electrons (see Fig.$\,$\ref{fig:echoZ}a), distort the valence orbitals and crystal lattice, thus creating EFGs at nuclear sites (see Fig.$\,$\ref{fig:echoZ}b). These EFGs couple to the quadrupolar momentum of the nuclei with spin~$I=3/2$ and modify the splitting of the $I_z =\pm3/2\leftrightarrow\pm1/2$ satellite Larmor transitions by \cite{sundfors_exchange_1969}
\begin{equation}\label{eq:quad}
	\omega_{\mathrm{Q},\alpha} =  \frac{eQ_\alpha}{2} V_{x'x'},
\end{equation}
where $Q_\alpha$ is the quadrupolar moment of nuclear species $\alpha$ and $V_{x'x'}$ denotes the component of the electric field gradient tensor in the direction of the external field (Fig\,\ref{fig:echoZ}b). The local field gradients are given by \cite{sundfors_exchange_1969}
\begin{equation}\label{eq:grad}
	V_{x'x'} = R_{14,\alpha} E_z \cos \left( 2\theta \right).	
\end{equation}
$R_{14,\alpha}$ is the species dependent response tensor component relating electric fields to electric field gradients at the nuclear site and $\theta$ is the angle between the magnetic field and
the [110] axis. The variation of the local electric field $E_z$ across the
electronic wave function due to the electron's own charge density introduces a broadening of the precession frequencies. The dependence of $\omega_{Q,\alpha}$ on $\theta$, arising from the crystal
symmetry of the host material, implies a suppression of the effect for a field along the [100] and [010] axis.

The Hahn echo amplitude as a function of separation time is shown in Fig.$\,$\ref{fig:echoY} for different in-plane field directions $\theta$ between the [110] and the [1$\bar{1}$0] axes. Indeed a factor two longer coherence is seen for $\theta = 45^{\circ}$, parallel to the [100] (or [010]) direction. Apart from this enhancement, another oscillatory modulation appears, reaching a maximum at the same angle.

\begin{figure}
	\centering
	\includegraphics[trim = 0cm 0cm 0cm 0cm, width=\textwidth/2]{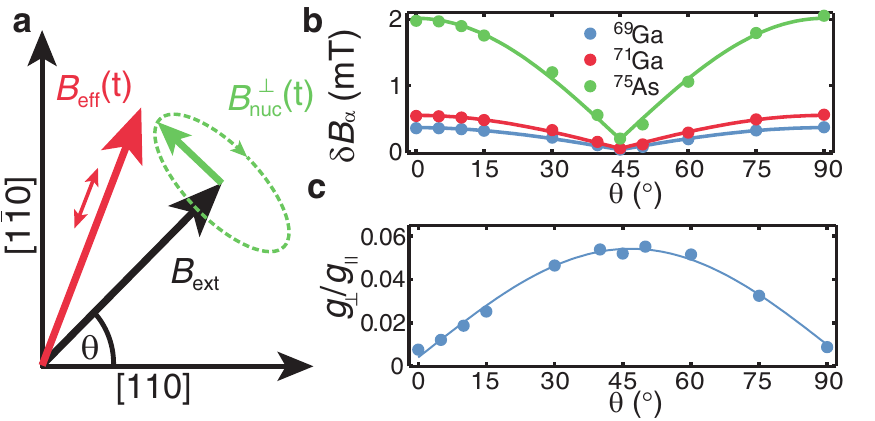}
	\caption{\textbf{\textit{g}-factor anisotropy and fit parameters} \textbf{a}, Due to an anisotropic $g$-tensor electron and nuclear spins have different quantization axes, $B_{\mathrm{eff}}$ and $B_{\mathrm{ext}}$, respectively. This leads to a linear contribution of the transverse Overhauser field to the electronic Zeeman splitting, oscillating with the Larmor frequencies of the nuclear spins. \textbf{b,\,c}, Fit parameters extracted for different in-plane magnetic field directions $\theta$. The quadrupolar contribution to  nuclear broadening $\delta B_\alpha = \hbar \delta\omega_{Q,\alpha} /\gamma_\alpha$, expressed in terms of an equivalent line width for the three isotopes using equation\,(\ref{eq:quad}), vanishes at 45$^\circ$ along the [100] direction. At the same angle, the $\sin(2\theta)$ dependence of the coupling $g_{\perp}$ to the transverse hyperfine field reaches a maximum. }
	\label{fig:params}	
\end{figure}

To further investigate the origin of these oscillations we aligned $B_\mathrm{ext}$ along the [100]-axis and varied its magnitude in Fig.$\,$\ref{fig:echoXZ}. With decreasing $B_\mathrm{ext}$ the frequency of the modulation decreases, until at $100\,$mT only a very fast decay of the echo amplitude followed by a revival at $\tau \approx 13\,\mu$s occurs. This envelope modulation can be explained by an electronic $g$-factor anisotropy, arising from an asymmetric confinement of the electron in the 2DEG and spin-orbit coupling\cite{v._k._kalevich_electron_1993, nefyodov_electron_2011, snelling_magnetic_1991}.  The main axes of the $g$-tensor are expected to be the [110] and [1$\bar{1}$0] crystal axis, consistent with the absence of a fast echo
modulation with $B$ along these directions.
For other field directions,  the quantization axis of the electron differs from the external field around which the nuclear spins precess. A linear coupling to the transverse nuclear magnetic field $B_{\mathrm{nuc}}^{\perp}$ thus appears in the effective magnetic field determining the electronic Zeeman splitting (see Fig.$\,$\ref{fig:params}a and Supplementary Information):
\begin{equation}\label{eq:bfield}
	B_{\mathrm{eff}} = g_\parallel B_{\mathrm{ext}} + g_{\perp} B_{\mathrm{nuc}}^{\perp}(t),
\end{equation}
where $g_{\parallel} (g_{\perp})$ denotes the \mbox{(off-)diagonal} entries of the $g$-tensor. 
During the free evolution part of the spin echo, the qubit acquires a phase 
arising from $B_{\mathrm{nuc}}^{\perp}(t)$. Due to the dynamics of $B_{\mathrm{nuc}}^{\perp}(t)$ that phase is not eliminated by the echo pulse and hence leads to dephasing. But whenever the evolution time $\tau/2$ is a multiple of all three Larmor frequencies, the net phase accumulated vanishes and the echo amplitude recovers. Partial recovery occurs if the evolution time only matches a multiple of the Larmor period of two or one species.

To obtain a quantitative description of quadrupolar and anisotropy effects, we adapt the semiclassical model of Ref.\,\onlinecite{bluhm_dephasing_2011}, based on computing the total electronic phase accumulated due to
the precessing nuclear spins and averaging over the initial nuclear state. The transverse hyperfine field is modeled as the vector sum of Gaussian distributed contributions arising from the three nuclear species and
the spread of quadrupolar shifts. The distribution of nuclear precession frequencies $F(\omega)$ is chosen such that the correlation function of the transverse field is that obtained from an ensemble of independent nuclear spins 3/2 subjected to a Gaussian distribution of quadrupolar shifts (see Supplementary Information).
$F(\omega)$ is taken as the weighted sum of two Gaussians centered on the Larmor frequency, reflecting the contributions from the unperturbed center transition and the quadrupole broadened satellite transitions as schematically depicted in Fig.$\,$\ref{fig:echoZ}d.
The rms-width of the quadrupolar broadened distribution is given by the variation of electric fields via equation\,(\ref{eq:quad}) and (\ref{eq:grad}).\newline
Using this model we fit the data (Fig.$\,$1-3) with most free parameters being independent of the magnetic field (see Supplementary Information).
Most relevant for this work are the quadrupolar broadenings of nuclear transition and the linear coupling to transverse hyperfine fields $g_{\perp}$ (both depending on field direction only) shown in Fig.$\,$\ref{fig:params}b and c. As predicted, the quadrupolar broadening approximately vanishes at $ \theta = 45^{\circ}$ and is maximal at $\theta = 0^{\circ}$ and $\theta = 90^{\circ}$.
The maximum magnitude of $\delta B_\alpha$ is consistent with the electric field variation generated by the electron in the dot (see Supplementary Information). The off-diagonal $g$-tensor element $g_{\perp}$ shows the predicted $\sin{(2\theta)}$ dependence, and its maximum anisotropy of $5\,\%$ is comparable with that found in quantum wells \cite{nefyodov_electron_2011}.

One of our key results is that quadrupole broadening of nuclear spins can contribute to electronic dephasing by decorrelating the transverse nuclear polarization, which contributes to the electronic Zeeman splitting to second order.
While in principle another source of anisotropy with the same angular dependence could explain the observed variation of the coherence time, we are not aware of any other plausible mechanism. Anisotropic diffusion\cite{witzel_wavefunction_2008} shows a different angular dependence with the longest coherence times along the $[110]$ direction. Our interpretation is further supported by the good quantitative agreement with the model and NMR measurements\cite{sundfors_exchange_1969,hester_nuclear-magnetic-resonance_1974}.
This result does not contradict the reported suppression of 
nuclear spin diffusion\cite{maletinsky_dynamics_2007} by quadrupole effect\cite{chekhovich_suppression_2015} as spin diffusion mostly affects electron coherence via the longitudinal polarization, whereas in our case the transverse coupling is dominant. An isotropic $g$-factor in combination with an anisotropic hyperfine interaction would lead to the same echo modulation when rotating $B_{\mathrm{ext}}$, but the anisotropy of the hyperfine interaction is usually assumed to be negligible as the conduction band wavefunction of GaAs is predominantly s-type.

While in the present sample $g$-factor anisotropy and quadrupolar effects cannot be eliminated simultaneously, symmetric, possibly back-gated quantum wells\cite{nefyodov_electron_2011} should allow the elimination of any $g$-factor anisotropy. The back gate could also be used to tune quadrupolar interaction, as it depends on the electric field, thus allowing further studies. 
Given that the strain-induced quadrupole broadening  in self-assembled dots was found to be 3-4 orders of magnitudes larger \cite{chekhovich_suppression_2015}, 
it likely also has pronounced effects on the coherence of this type of quantum dot, which is currently less well understood than that of gated dots.
Furthermore, the echo modulation due to an anisotropic $g$-factor may also play an important role in III-V nanowire qubits, where strong $g$-factor anisotropies and short coherence times have been measured\cite{nadj-perge_spectroscopy_2012, nadj-perge_spin-orbit_2010}. 
\section{Methods}

\subsection{Qubit system.}

The quantum dots used in this work were fabricated on a GaAs/Al$_{0.69}$Ga$_{0.31}$As heterostructure with Si-$\delta$-doping $50\,$nm below the surface and a spacer thickness of $40\,$nm, leaving the 2DEG at 90\,nm depth, as shown in Fig.$\,$\ref{fig:echoZ}a (see Supplementary Information).

\subsection{Echo sequence.}

Following the experimental procedure for Hahn spin echo measurements from Ref.~\onlinecite{bluhm_dephasing_2011} we first initialize the qubit system in the spin singlet groundstate $S$ by pulsing both electrons into one dot. Rapidly separating the electrons into both dots lets them evolve in different Zeeman fields arising from the external magnetic field $B_{\mathrm{ext}}$ and the fluctuating local Overhauser field $B_{\mathrm{L(R)}}$ for a time $\tau$. A gradient $\Delta B_{z}=|B_{\mathrm{L}}-B_{\mathrm{R}}|/2$ in the hyperfine field of the two dots leads to coherent rotations between $S$ and $T_{0}$ and fluctuations in $\Delta B_{z}$ cause dephasing. An exchange splitting between the spin singlet $S$ and triplet state $T_{0}$ arises from inter-dot tunnel-coupling. This exchange allows electric control of the qubit by varying the difference in electrostatic potential between the two dots, on the nanosecond timescale with an arbitrary waveform generator.  Using this exchange interaction to perform a $\pi$-pulse by driving rotations between the eigenstates $\ket{\uparrow\downarrow}$ and $\ket{\downarrow\uparrow}$, we swap the two electrons halfway through the evolution time $\tau$. Lastly, we read out the final qubit state by pulsing the electrons into one dot. Using Pauli-spin-blockade we distinguish between singlet and triplet states by measuring the resistance of a nearby sensing dot via RF-reflectometry. Such a pulse cycle with varying evolution times is repeated several million times and the average echo amplitude is recorded. Simultaneous histogramming of individual measurement outcomes is used for normalization\cite{barthel_rapid_2009}.
The fine tuning of the pulses that was necessary in Ref.\,\onlinecite{bluhm_dephasing_2011} to avoid artifacts from  shifts of the wave function has been eliminated due to improved RF-engineering. 

\subsection{Acknowledgements}
This work was supported by the Alfried Krupp von Bohlen und Halbach Foundation and DFG grant BL 1197/2-1 and SFB 689.

\subsection{Author contribution}
Molecular-beam-epitaxy growth of the sample was carried out by
D.S. and D.B.. T.B. and R.P.G.M. set-up the experiment. T.B. fabricated the sample and conducted the experiment. T.B. and H.B. developed the theoretical model, analyzed the data and wrote the paper.

\subsection{Additional information} 
The authors declare no competing financial interests. Online supplementary information
accompanies this paper. 
Correspondence and requests for materials should be addressed to H.B.

\bibliographystyle{naturemag}

\end{document}